\documentclass{article}

\usepackage{arxiv}
\usepackage{array} 
 \newcolumntype{L}{>{\centering\let\newline\\\arraybackslash\hspace{0pt}}m{2cm}}
\usepackage{threeparttable} 
\usepackage{multirow}
\usepackage{epsfig}

\usepackage{amsmath}
\usepackage[utf8]{inputenc} 
\usepackage[T1]{fontenc}    
\usepackage{hyperref}       
\usepackage{url}            
\usepackage{booktabs}       
\usepackage{amsfonts}       
\usepackage{nicefrac}       
\usepackage{microtype}      
\usepackage{lipsum}
\usepackage{graphicx}
\graphicspath{ {./images/} }

\title{A Precision Trial Case Study for Heterogeneous Treatment Effects in Obstructive Sleep Apnea}

\author{
 Lara Maleyeff \\
 Department of Epidemiology, Biostatistics, and Occupational Health \\ McGill University, Montr\'eal, QC, CA \\
  \texttt{lara.maleyeff@mcgill.ca} \\
   \And
 Shirin Golchi \\
 Department of Epidemiology, Biostatistics, and Occupational Health \\ McGill University, Montr\'eal, QC, CA \\
  \And
 Erica E. M. Moodie \\
Department of Epidemiology, Biostatistics, and Occupational Health \\ McGill University, Montr\'eal, QC, CA \\
\And
R. John Kimoff \\
Respiratory Division and Sleep Laboratory \\ McGill University Health Centre, Montréal, QC, Canada \\ \\
Translational Research in Respiratory Diseases Program \\ 
Research Institute of the McGill University Health Centre, Montréal, QC, Canada \\
}

\begin{document}
\maketitle
\begin{abstract}
Precision medicine tailors treatments to individual patient characteristics, which is especially valuable for conditions like obstructive sleep apnea (OSA), where treatment responses vary widely. Traditional trials often overlook subgroup differences, leading to suboptimal recommendations. Current approaches rely on pre-specified thresholds with inherent uncertainty, assuming these thresholds are correct—a flawed assumption. {This case study} compares pre-specified thresholds to two advanced Bayesian methods: the established FK-BMA method and its novel variant, FK. The FK approach retains the flexibility of free-knot splines but omits variable selection, providing stable, interpretable models. Using biomarker data from large studies, this design identifies subgroups dynamically, allowing early trial termination or enrollment adjustments. {Simulations in this specific context} show FK improves precision, efficiency, and subgroup detection, offering practical benefits over FK-BMA and advancing precision medicine for OSA.
\end{abstract}



\section{Introduction}

Precision medicine, which tailors medical interventions based on patient characteristics, can offer several advantages compared to a one-size-fits-all approach, including improved treatment efficacy, reduced adverse effects, optimized resource allocation, and ultimately, better patient outcomes. This approach is particularly impactful in conditions like obstructive sleep apnea (OSA) where treatment responses vary widely among individuals. In such settings, the treatment that is considered optimal for a patient with one set of characteristics might not be ideal for another. Traditional clinical trials are often designed to detect population-averaged effects, which may overlook significant differences in treatment efficacy among subgroups of patients. In light of these challenges, individualized, evidence-based clinical decision-making strategies that account for such heterogeneity are increasingly favored and are gaining  popularity in medical research \cite{thall2021adaptive,baldi2023new,fisher2018stochastic,lai2019adaptive}.

Bayesian adaptive enrichment clinical trials adjust enrollment based on interim data, making them more efficient than traditional randomized controlled trials (RCTs) in detecting subgroup differences. They allow for continuous updating of evidence, enabling faster decision-making about the effectiveness of a biomarker. This approach reduces both the time and resources needed to identify potential therapeutic targets or diagnostic markers. These innovative designs and their various adaptations \cite{wang2007approaches, magnusson2013group,rosenblum2016group, brannath2009confirmatory, jenkins2011adaptive,mehta2014biomarker}, as well as basket trials, are gaining popularity for their cost-effective assessment of a single treatment's impact across different patient groups identified by specific biomarkers \cite{asano2020bayesian,chen2020bayesian, chu2018bayesian, jin2020bayesian, park2020overview,simon2017critical}. Basket trials are a a family of designs most commonly used in oncology that test the effect of one drug on a single mutation in a variety of tumor types, allowing for a more streamlined and efficient evaluation of the treatment’s effectiveness across multiple patient populations. Recent developments in this domain include a  basket trial design that uses Bayesian model averaging for enhanced decision-making by borrowing information across similar groups \cite{psioda2021bayesian}.

The focus of most biomarker-driven trial designs is on predefined patient subgroups, categorized by a fixed set of biomarkers. However, innovative designs also facilitate the identification of biomarkers during the trial by, for example, using machine learning or Bayesian lasso for biomarker refinement during interim analyses as previously proposed \cite{freidlin2005adaptive, freidlin2010cross}. These approaches, including the Bayesian group sequential enrichment design \cite{park2022bayesian} which uses spike-and-slab priors as a means of implicit variable selection, aim to streamline biomarker selection and enhance treatment assessment within responsive subgroups. Meanwhile, advancements in identifying optimal biomarker thresholds have evolved to accommodate nonlinear and nonmonotone relationships between biomarkers and treatment effects, with \cite{liu2022bayesian} proposing a design that directly estimates these complex relationships, demonstrating improvements over traditional methods. More recently, \cite{maleyeff2024adaptive} developed FK-BMA (Free-Knot Bayesian Model Averaging), a Bayesian adaptive enrichment design which identifies important predictive variables out of a larger set of candidate biomarkers. They use a flexible modeling framework that incorporates the effects of continuous biomarkers through free-knot B-splines and estimate parameters by marginalizing over all possible variable combinations using Bayesian model averaging. Interim analyses evaluate whether a biomarker-defined subgroup has altered treatment effects, allowing for early termination due to efficacy or futility and limiting future enrollment to treatment-sensitive patients.

OSA is a highly prevalent condition affecting an estimated 35\% of Canadian adults and is characterized by repeated episodes of upper airway obstruction during sleep. Epidemiologic studies have established OSA as an independent cardiovascular (CV) risk factor, with evidence linking OSA to adverse CV outcomes such as hypertension, myocardial infarction, stroke, arrhythmias, and increased CV mortality \cite{benjafield2019estimation,cowie2021sleep}. Non-randomized interventional studies have demonstrated significant CV benefits from positive airway pressure (PAP) treatment of OSA, with reductions in CV events and overall mortality \cite{patil2019treatment}. However, three recent prospective RCTs have shown neutral effects of PAP treatment on secondary CV prevention, raising questions about the potential CV benefits of OSA treatment \cite{mcevoy2016cpap, peker2016effect,sanchez2020effect}.

One key factor contributing to these discrepant findings may be participant heterogeneity, as there are substantial differences in individual susceptibility to the CV sequelae of OSA and the potential benefits of OSA treatment. Recent work has identified two physiologic biomarkers—mean hypoxic burden (HB) and heart rate response to apneas/hypopneas ($\Delta$HR)—which are derived from routine sleep recordings and have shown predictive value for adverse CV outcomes in retrospective analyses \cite{azarbarzin2019hypoxic,azarbarzin2021sleep}. In addition to these established biomarkers, vasoconstrictive burden (VCB) has emerged as another promising metric \cite{hajipour2024all}.

{

Motivated by the challenges of patient heterogeneity in OSA, we provide a case study demonstrating how advanced Bayesian adaptive enrichment clinical trial designs can be applied to identify whether high-risk patients, based on biomarkers HB and $\Delta$HR, experience substantial reductions in systolic blood pressure during a 6-month PAP treatment. The work introduces the free-knot (FK) method as a novel approach inspired by the FK-BMA framework proposed by \cite{maleyeff2024adaptive}. Unlike FK-BMA, the FK method excludes variable selection, focusing on a small number of pre-specified biomarkers based on prior evidence. While the FK method offers flexibility for modeling complex, nonlinear biomarker relationships, this study emphasizes its utility as part of a broader framework for practical implementation rather than as a definitive solution. Furthermore, this is the first study to apply these advanced Bayesian methods in a real-world context, using biomarker distributions derived from the Sleep Heart Health Study (SHHS) and the Multi-Ethnic Study of Atherosclerosis (MESA). 

}

\section{Motivating setting}
\label{sec:trial_setup}

Our motivating example is a prospective, 6-month, single-blind RCT designed to evaluate the effects of PAP treatment on CV risk markers in patients with moderate-severe OSA. This study employs a Bayesian adaptive enrichment clinical trial design to identify subgroups, defined by continuous biomarkers HB and $\Delta$HR, where PAP treatment is hypothesized to result in improvements in 24-hour systolic blood pressure. 

We selected systolic blood pressure as the primary outcome because it is a well-established predictor of future CV events and overall mortality, with elevated systolic blood pressure directly linked to an increased risk of major adverse CV events such as heart attacks and strokes \cite{trialists2005effects,lewington2002age}. Changes in systolic blood pressure provide a sensitive and meaningful measure of CV health and potential benefits from interventions aimed at reducing CV risk. {We focus on effect sizes of 5 to 7.5 mmHg and an individual-level standard deviation of 8.4 mmHg for this study based on meta-analyses of PAP treatment in OSA. These correspond to signal-to-noise ratios of 0.60 and 0.89 and \( R^2 \) values of 0.37 and 0.47, respectively.
\cite{patil2019treatment}.}

Participants in the treatment group will receive PAP therapy. The intervention will include an educational session on healthy sleep habits, OSA, and PAP treatment, including mask fitting and hands-on training. The control group will receive conservative therapy, including healthy sleep counseling and nasal dilator strips (NDS) to use during sleep. NDS may reduce snoring and improve subjective sleep quality but do not impact OSA severity or outcomes \cite{amaro2012use}.

Study measurements will be obtained at baseline and repeated at 6 months. These will include anthropometric data, medical history, and medication records, as well as 24-hour systolic blood pressure. HB will be calculated as the total area under the oxygen desaturation curve divided by sleep time, and $\Delta$HR will measure heart rate increases linked to apneas and hypopneas. These measurements will help assess the impact of PAP therapy on cardiovascular risk in OSA patients. VCB will be measured using a finger sensor during sleep, tracking pulse signals where drops in amplitude indicate vasoconstriction. 

\section{Methods}
\label{sec:methods}

\subsection{Notation}
\label{sec:notation}

For patient $i$, let $Y_i$ represent the negative observed systolic blood pressure (such that positive effects represent decreases), $T_i$ denote the assigned treatment indicator ($T_i=0$ for control, $T_i=1$ for treatment), $x_{1i}$ be the HB, and $x_{2i}$ be the $\Delta$HR. 

\subsection{Trial hypothesis}
\label{sec:trial_hypothesis}

In order to detect heterogeneous treatment effects, a common goal is to identify the effective subspace. First, we let $\gamma(\mathbf{x})$ be the variable-specific treatment effect or the blip effect of the treatment \cite{robins1997causal}. Then the effective subspace can be defined as
\begin{equation}
\label{eq:eff_subspace}
    \mathcal{X}^* = \left\{\mathbf{{x}}: P\left( \gamma(\mathbf{x}) > 0 \mid \mathcal{D}\right)>1-\alpha\right\},
\end{equation}
where $\alpha$ is specified through simulation, $\mathbf{x}=(x_1,x_2)$, and $\mathcal{D}$ is the observed data. For example, $\mathcal{X}^*$ can consist of HB $>$ 60 (\%min)/h and $\Delta$HR $>$ 8 bpm, implying that patients with those characteristics are expected to benefit from the treatment. Given the effective subspace, the hypothesis of the trial designed to detect a treatment effect in the effective subspace can be written as 
\begin{equation}
\begin{split}
      H_0&: \gamma(\mathbf{x}) \le 0 \text{ } \forall \text{ }\mathbf{{x}} \in \mathcal{X}  \text{ vs. } \\
      H_A&: \exists \text{ } \mathbf{{x}} \in \mathcal{X}  \text{ s.t. } \gamma(\mathbf{x}) > 0,
\end{split}
\end{equation}
where $\mathcal{X}$ is the complete space of candidate predictive variables. In other words, the null hypothesis states that the treatment is ineffective for all values of HB and $\Delta$HR and the alternative hypothesis is that there exists at least one variable combination for which the treatment is effective. 

\subsection{Model formulations}

We compare three model formulations: a model based on a pre-specified threshold for continuous biomarkers HB and $\Delta$HR, which we refer to as the Cutoff model; the Free Knot-Bayesian Model Averaging (FK-BMA) method of \cite{maleyeff2024adaptive}, a Bayesian model averaging approach that automatically identifies optimal cutoffs and incorporates variable selection to identify the most relevant predictors; and the Free Knot (FK) model, similar to FK-BMA but without any variable selection, treating all biomarkers as equally important predictors.

The pre-specified cutoffs for HB $>$ 60 (\%min)/h and $\Delta$HR $>$ 8 bpm were selected based on retrospective analyses demonstrating their predictive value for increased CV risk and morbidity in OSA patients \cite{azarbarzin2019hypoxic,azarbarzin2021sleep}. These evidence-based thresholds are designed to improve the precision of patient selection by focusing on those most likely to benefit from PAP therapy. However, it is important to note that these thresholds are based on estimates, and the uncertainty surrounding them is not accounted for. Specifically, the Cutoff model formulation is as follows. 
\begin{equation}
\label{eq:complete}
    Y_i = \beta_0 + \beta_1z_{1i} + \beta_2z_{2i} + \beta_3z_{1i}z_{2i} + \left(\beta_4 + \beta_5z_{1i} + \beta_6z_{2i} + \beta_7z_{1i}z_{2i}\right) T_i + \epsilon_i,
\end{equation}

where $z_{1i}=I(x_{1i}>60)$, $z_{2i}=I(x_{2i}>8)$, and $\epsilon_i \sim \mathcal{N}(0,\sigma_\epsilon^2)$. Here, $\gamma(\mathbf{x})=\beta_4 + \beta_5z_{1} + \beta_6z_{2} + \beta_7z_{1}z_{2}$. We postulate the following non-informative prior distributions:
\begin{equation}
\begin{split}
    \beta_k &\sim \mathcal{N}(0,20) \text{ for } k=1,\dots,6\\
    \sigma_\epsilon^2 &\sim \text{\textit{Inverse-Gamma}}(0.01,0.01)
\end{split}
\end{equation}

For the FK-BMA and FK formulations, the saturated model can be written as:
\begin{equation}
\label{eq:complete3}
    Y_i= \sum_{j=1}^{2} h_{j}(x_{ji}) + h_{12}(x_{1i}x_{2i}) + \left(\phi + \sum_{j=1}^{2} f_{j}(x_{ji}) + f_{12}(x_{1i}x_{2i})\right) T_i + \epsilon_i,
\end{equation}
where $h_{1}$, $h_{2}$, $h_{12}$, $f_{1}$, $f_{2}$, and $f_{12}$ are biomarker terms represented by free-knot B-spline functions with degree 1 and $\epsilon_i \sim \mathcal{N}(0,\sigma_\epsilon^2)$. Here, the variable-specific treatment effect is $\gamma(\mathbf{x}) = \phi + \sum_{j=1}^{2} f_{j}(x_{j}) + f_{12}(x_{1}x_{2})$. { We opted to use linear B-splines rather than cubic B-splines as were used in \cite{maleyeff2024adaptive} in the presence of a lower signal-to-noise ratio as (1) the constraints of the trial did not permit the fitting of more complex cubic B-splines, and (2) the primary goal of the trial was to identify accurate biomarker cutoffs, rendering the finer granularity captured by cubic B-splines less critical as long as the cutoff was correctly identified. To assess the impact of fitting linear splines to non-linear functions, see Scenarios 4-5 and 7-8 in the simulation study.} For both FK-BMA and FK, we use non-informative priors for both the model coefficients, $\mathcal{N}(0,20)$, and the individual heterogeneity, $\sigma_\epsilon^2 \sim \text{\textit{Inverse-Gamma}}(0.01,0.01)$. Additionally, we assign a truncated Poisson prior distribution to the number of knots in each spline term, $k_s$. Thus,

\begin{equation}
    p(k_s \mid \lambda_2) = \frac{e^{-\lambda_2}\lambda_2^{k_s}}{C_2 k_s!},
\end{equation}
where $C_2$ is a normalization constant. All knot locations from a set of candidate internal knots of size $K_s$ have equal probability of inclusion. The candidate internal knots are placed at the $(q/6)$-quantiles ($q=1,\dots,5$) for each biomarker, ensuring data-adaptive results while maintaining practical feasibility.

For the FK-BMA model only, we additionally assign a prior distribution on which biomarker terms ($h_{1}$, $h_{2}$, $h_{12}$, $f_{1}$, $f_{2}$, and $f_{12}$) are included in the model. This facilities Bayesian model averaging, which involves averaging over multiple candidate models weighted by their posterior probabilities to account for model uncertainty. Candidate models are identified based on different combinations of biomarker terms. Here, we postulate a truncated Poisson prior distribution on the number of biomarker terms in the model $m$:
\begin{equation}
    p(m \mid \lambda_1) = \frac{e^{-\lambda_1}\lambda_1^m}{C_1 m!},
\end{equation}
where $C_1$ is a normalization constant. We assume that all biomarker terms have equal probability of being included in the model. Both $\lambda_1$ and $\lambda_2$ must be specified \textit{a priori}; larger $\lambda_1$ favors larger models and $\lambda_2$ favors  more complex splines. We consider $\lambda_1=\{2,3\}$ and $\lambda_2=3$ in the simulation study. All prior parameters were selected to be non-informative, with the exception of $\lambda_1$ and $\lambda_2$, which are tuned by simulation.

Standard computational methods cannot be used for fitting the FK-BMA and FK models due to the need to navigate complex, high-dimensional model spaces with varying dimensions. These models involve both the selection of spline knots and biomarker terms (FK-BMA only), for which the reversible jump Markov Chain Monte Carlo (rjMCMC) sampler is a highly suitable approach \cite{green2009reversible,maleyeff2024adaptive}. This algorithm efficiently explores the model space by proposing changes to spline knots and biomarker terms, ensuring model hierarchy is preserved. {Here, model hierarchy  ensures that if an interaction effect between a biomarker and treatment is included in the model, the corresponding main effects of that interaction are also included. In our framework, we treat the interaction between HB and $\Delta$HR as a distinct biomarker. As such, the inclusion of the HB-$\Delta$HR interaction with treatment in the model guarantees the presence of the main effect of HB-$\Delta$HR, but it does not imply that the individual main effects of HB and $\Delta$HR are necessarily included in the model. The main effect of treatment is always included in the model to ensure consistent interpretation of treatment effects and uphold the clinical relevance and comparability of results across models in the trial.} The rjMCMC procedure dynamically transitions between different dimensional spaces by adding or removing knots and terms, adjusting associated coefficients accordingly. These steps generate a sequence of states representing the posterior distribution, allowing for flexible modeling of continuous biomarkers and averaging over possible models, thereby integrating model uncertainty and facilitating robust inference in Bayesian adaptive enrichment designs.

\subsection{Interim analysis and enrichment}
\label{sec:interim}

We employ the interim analysis procedure of \cite{maleyeff2024adaptive}, with a slight modification. As before, decisions are made based on the posterior distribution of the treatment effect averaged over the enriched population. At each interim analysis, we identify the ``effective subspace" using the accumulated data. However, if the prevalence of this subspace is outside a predefined threshold range (0.05, 0.95), we widen the effective subspace to include the entire sample and perform a non-subgroup-specific analysis, effectively reverting to an overall trial population assessment. This adjustment ensures the trial adapts dynamically to include the full population if subgroup differentiation becomes irrelevant. The trial can be stopped early for efficacy or futility based on posterior probabilities in both the subgroup and overall population, improving efficiency while maintaining flexibility. We use the efficacy and futility cutoffs, $B_1$ and $B_2$ respectively, as defined previously \cite{maleyeff2024adaptive}.

If the trial is not stopped early, at the last interim analysis, we conclude that the treatment is effective in the sensitive population using the efficacy criterion. Otherwise, we conclude that treatment is not superior in the sensitive population. This design accounts for the possibility that the entire trial population is in the effective subspace. In this case, we assess the average treatment effect in the entire trial sample, as in a traditional randomized clinical trial.

For FK-BMA, we incorporate an additional variable selection step at the end of the trial. FK-BMA does not explicitly remove irrelevant variables during sampling; instead, these variables are excluded from most of the models in the rjMCMC sampler and have a low posterior inclusion probability. We propose pruning variables with a posterior inclusion probability of less than 10\%. In practice, such simplifications can ease clinical implementation and reduce costs.

\section{Simulation Study}
\subsection{Simulation set-up}
\label{s:sim_setup}
In this section, we summarize the results of a simulation study to assess the operating characteristics of three trial designs that differ in their model formulation: the Cutoff, FK-BMA, and FK methods. We used the fitting procedures from the respective modelling formulations, with the interim decision rules described in the interim analysis and enrichment procedure. We assumed a maximum sample size of 500 with one interim analysis at 300 patients and considered HB, $\Delta$HR, and their interaction as candidate biomarkers. In order to define the effective subspace, we set various subsets of the variables to have a non-zero interaction with treatment; we refer to biomarkers with a non-zero interaction effect as the true predictive variables.
\begin{figure}
    \centering
      \caption{Histogram of 2000 samples from simulated predictive variables.}
    \includegraphics[width=0.5\linewidth]{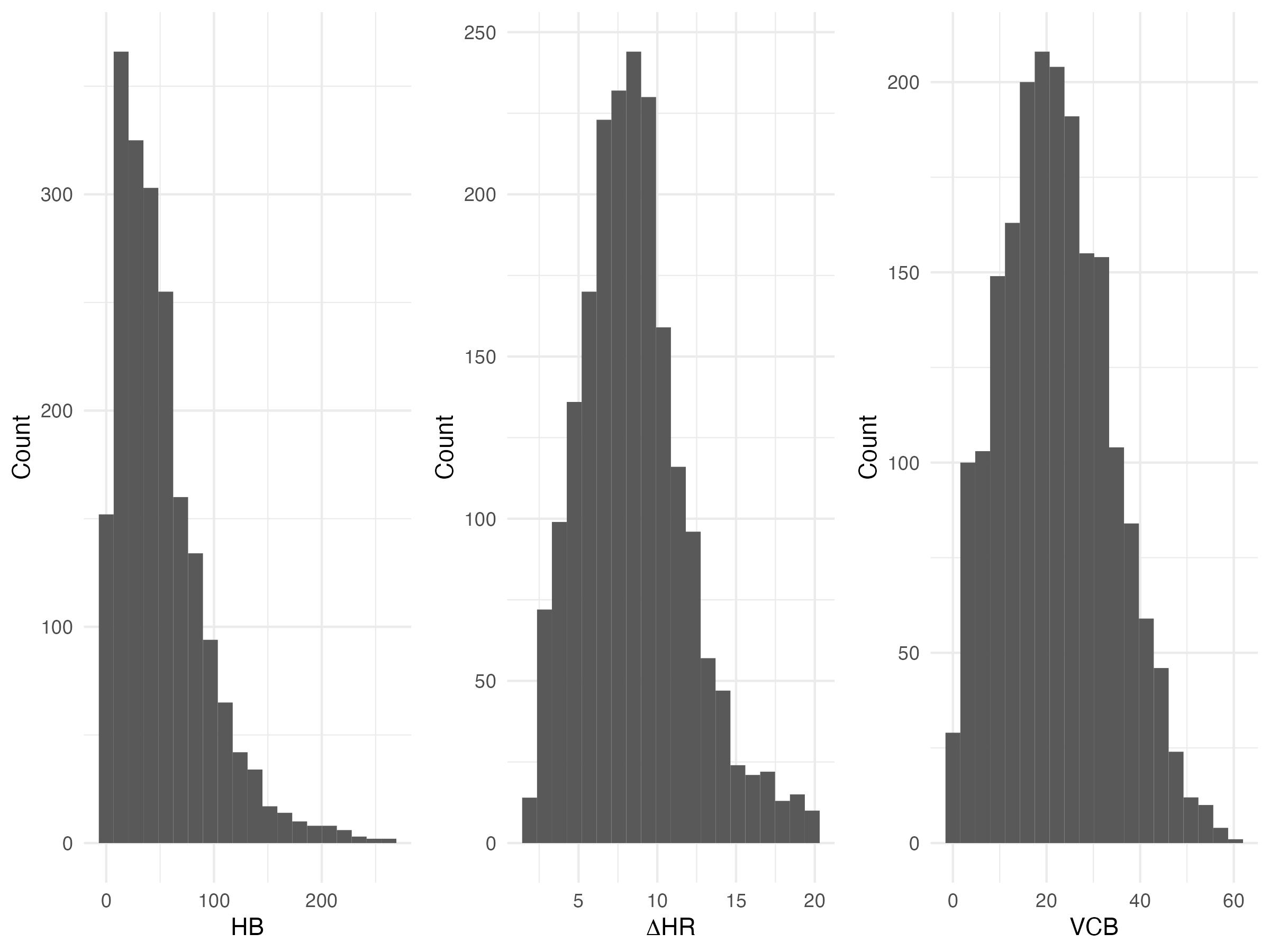}
    \label{fig:hist}
\end{figure}
We assessed eight simulation scenarios for each design, described in Table \ref{tab:model_params} and Figure \ref{fig:plot_scens}. Data were generated from 
\begin{equation}
    Y_i = 3 I(x_{1i} \ge 60) + \gamma(\mathbf{x}_i)T_i + \epsilon_i,
\end{equation}
where $\epsilon_i \sim N(0,8.4^2)$, based on conservative estimates from previous meta-analyses \cite{patil2019treatment}. { Values for HB ($x_1$) and $\Delta$HR ($x_2$) were generated based on their sample distribution from the SHHS ($n=5,111$) and the MESA ($n=1,395$), respectively \cite{azarbarzin2019hypoxic,azarbarzin2021sleep}, with degrees of freedom, scale, and shift iteratively tuned based on reported quantiles. HB values were generated from a truncated t-distribution with 5 degrees of freedom, scaled by 50 and shifted by 15, and doubly truncated at 0 and 265. Similarly, $\Delta$HR values were generated from a truncated t-distribution with 5 degrees of freedom, scaled by 3 and shifted by 8, and doubly truncated at 2 and 20 (Figure \ref{fig:hist}, Table \ref{tab:quantile_comparison}).} Based on design simulation studies, we selected parameters $B_1 = 0.98$, $B_2 = 0.8$, and $\alpha=0.05$ to maintain a type I error rate of $0.05$ in the FK method. We simulated each trial 1000 times per scenario.

\begin{table}[ht]
\centering
\caption{Comparison of reported quantiles between study data and simulated data ($n=10,000$).}
\label{tab:quantile_comparison}
\begin{tabular}{@{}lcc@{}}
\toprule
\textbf{Quantile} & \textbf{Observed (Study)} & \textbf{Simulated} \\ 
\midrule
Q1 & $<16$ (SHHS) & $<17$  \\
Q2 & $16 - 28$ (SHHS) & $17 - 33$  \\
Q3 & $28 - 43$ (SHHS) & $33 - 52$ \\
Q4 & $43 - 71$ (SHHS) & $52 - 79$ \\
Q5 & $>71$ (SHHS) & $>79$ \\
\hline
Q1 & 2-5.8 (MESA) & $2 - 6.3$ \\
Q2-Q3 & 5.8-10.1 (MESA) & $6.3 - 10.4$ \\
Q4 & 10.1-20 (MESA) & $10.4 - 20$ \\ \bottomrule
\end{tabular}
\end{table}

\begin{table*}[!ht]
\centering
\caption{True model parameters used for Simulation Study. $\Delta$ is the true average treatment effect in the effective subspace. $F_{x_1} = \text{ecdf}(x_1)(x_1)$, $F_{x_2} = \text{ecdf}(x_2)(x_2)$, where ecdf is the empirical cumulative distribution function.}
\label{tab:model_params}
\begin{tabular*}{\textwidth}{cLcLcc}
\toprule
& & \multicolumn{4}{c}{Effective Subspace} \\
Scenario & Predictive Variables & $\gamma(\mathbf{x})$  & Correct Cutoff & Prevalence & $\Delta$ \\
\hline 
    1  & None & 0 & - & 0 & 0 \\
    2 & HB and $\Delta$HR & $5I(x_1>60,x_2>8)$& Yes & 0.18 & 5.0\\
    3  & HB & $5I(x_1>30,x_1<100)$  & No  & 0.52 & 5.0\\
    4 & HB & $\left[\frac{\exp\left\{30  \left(F_{x_1} - 0.5\right)\right\}}{1 + \exp\left\{30  \left(F_{x_1} - 0.5\right)\right\}}\right]  6.5 - 0.5$ &No &
0.58 & 5.0 \\    
5 & HB & $\left[
\begin{array}{ll}
\frac{\exp\{30  (F_{x_1} - 0.2)\}}{1 + \exp\{30  (F_{x_1} - 0.2)\}} & \text{if } F_{x_1} \leq 0.5, \\
\frac{1}{1 + \exp\{30  (F_{x_1} - 0.8)\}} & \text{if } F_{x_21} > 0.5
\end{array}
\right]  7.5 - 1$
 & No & 0.73 & 5.1\\

    6  & $\Delta$HR & $5I(x_2>8)$ & Yes & 0.54 & 5.0 \\
    7  & $\Delta$HR & $7.5I(x_2>12)$ & No & 0.15 & 7.5\\ 
    8  & $\Delta$HR & $\left[
\begin{array}{ll}
\frac{1}{1 + \exp\left\{100  \left(F_{x_2} - 0.2\right)\right\}} & \text{if } F_{x_2} \leq 0.5, \\
\frac{\exp\left\{100  \left(F_{x_2} - 0.8\right)\right\}}{1 + \exp\left\{100  \left(F_{x_2} - 0.8\right)\right\}} & \text{if } F_{x_2} > 0.5
\end{array}
\right]  6.5 - 0.5$ & No & 0.45 & 5.3\\ 
    \hline 
\end{tabular*}
\end{table*}

\begin{figure*}
    \centering
        \caption{Visual representation of Scenarios 4, 5, and 8.}
        \includegraphics[width=0.7\linewidth]{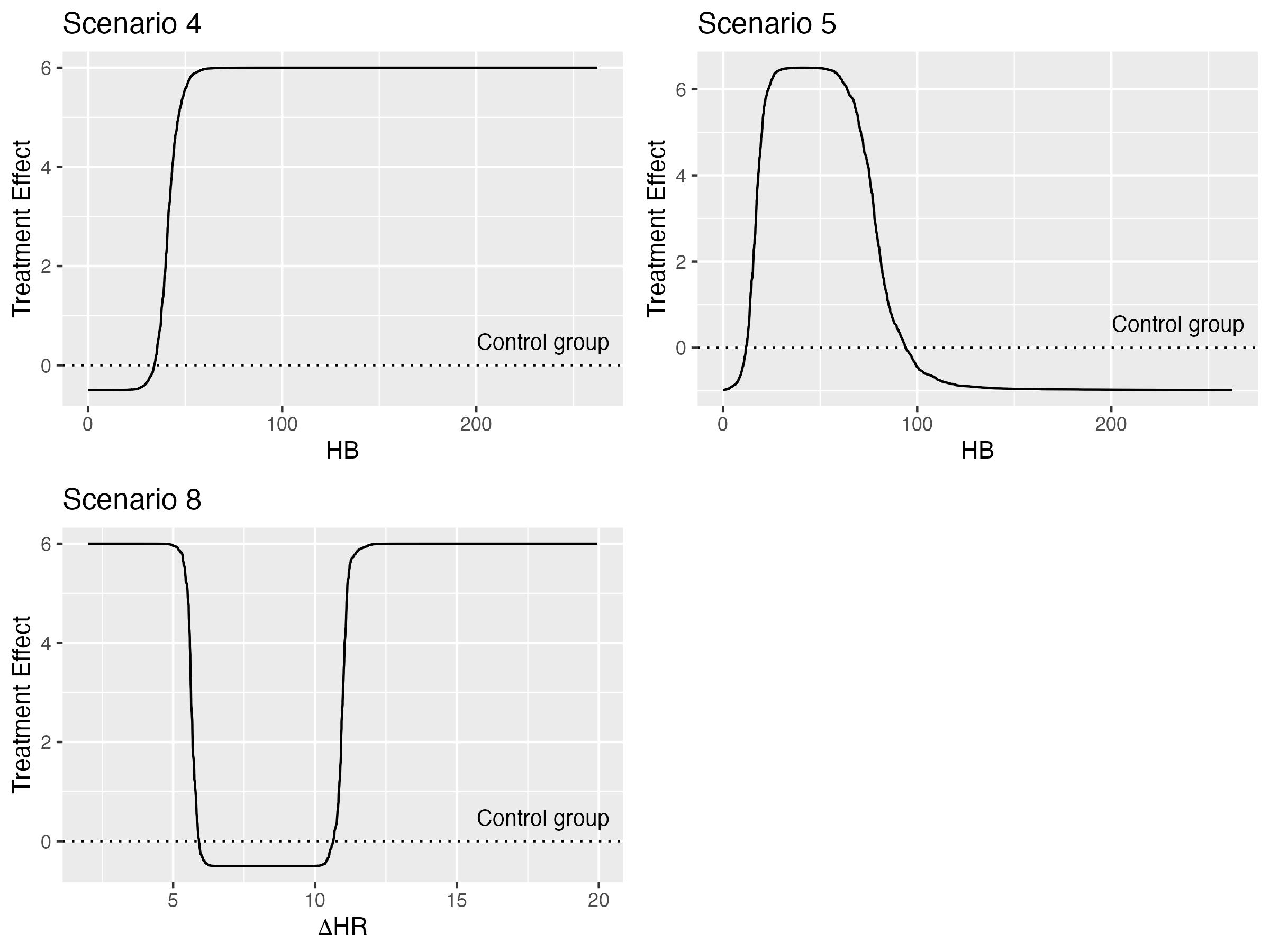}

    \label{fig:plot_scens}
\end{figure*}

As operating characteristics, we considered traditional type I error, power, exact detection rate (the success rate of exactly identifying the appropriate predictive variables without including any incorrect variables), inclusive detection rate (the success rate of having the appropriate predictive variables included in the detected set, even if other variables are also included), and accuracy of treatment recommendations using a large, external testing dataset. To assess accuracy of treatment recommendations, we generated an external dataset with 10,000 individuals and their biomarker information at the start of the simulation study. In each simulation iteration, we used the simulated trial results to recommend a treatment plan for individuals in the external dataset. Since the true functional relationship between biomarkers and the effect of treatment is known, we then calculated the percentage of individuals for which this treatment recommendation was optimal.

\subsection{Simulation results}
\label{s:sim_results}
\begin{table*}[t]
\centering
\caption{Empirical type I error (T1E; Scenario 1), power (Scenarios 2-8) and accuracy of treatment recommendations on an external dataset ($n=10,000$) for eight simulation scenarios.}
\label{tab:power_accuracy}
\begin{tabular}{c|cLLccLLc}
\toprule
  & \multicolumn{4}{c}{Power} & \multicolumn{4}{c}{Accuracy} \\ 
Scenario & Cutoff & FK-BMA ($\lambda_2 = 2$) & FK-BMA ($\lambda_2 = 3$) & FK & Cutoff & FK-BMA ($\lambda_2 = 2$) & FK-BMA ($\lambda_2 = 3$) & FK \\ 
  \hline
1 & 0.10 (T1E) & 0.04 (T1E) & 0.04 (T1E) & 0.08 & 0.96 & 0.96 & 0.97 & 0.96 \\ 
  2 & 0.80 & 0.26 & 0.28 & 0.58 & 0.92 & 0.71 & 0.71 & 0.78 \\ 
  3 & 0.91 & 0.89 & 0.91 & 0.97 & 0.51 & 0.57 & 0.61 & 0.73 \\ 
  4 & 1.00 & 0.96 & 0.97 & 1.00 & 0.69 & 0.75 & 0.77 & 0.82 \\ 
  5 & 1.00 & 0.99 & 0.99 & 1.00 & 0.64 & 0.75 & 0.77 & 0.83 \\ 
  6 & 0.99 & 0.91 & 0.93 & 0.99 & 0.92 & 0.59 & 0.62 & 0.73 \\ 
  7 & 0.49 & 0.35 & 0.41 & 0.68 & 0.75 & 0.69 & 0.70 & 0.75 \\ 
  8 & 0.76 & 0.75 & 0.77 & 0.89 & 0.51 & 0.54 & 0.57 & 0.66 \\ 
   \hline
\end{tabular}
\end{table*}

The empirical type I error, power, accuracy, predictive variable identification rates, and expected sample size of the methods were assessed across eight simulation scenarios. Scenario 1 represents the null scenario, where there is no treatment effect, and thus the power reflects the type I error rate. Scenarios 2 through 8 assess operating characteristics under various conditions, such as which variables are predictive, correct specification of cutoff values, and varying prevalence and effect sizes within the effective subspace. 

Table \ref{tab:power_accuracy} shows the power and accuracy metrics for the Cutoff, FK-BMA with two different model sparsity parameters, and FK methods. In Scenario 1, the type I error rate was well-controlled across the FK and FK-BMA methods, but inflated for the Cutoff method (0.10). When the cutoff was correctly specified, as in Scenarios 2 and 6, the Cutoff method performed well in terms of power, achieving 0.80 and 0.99, respectively. However, the FK method, although generally robust, showed lower power in Scenario 2 (0.58) when the prevalence of the effective subspace was low. Despite this, the FK method maintained adequate performance, particularly in Scenario 6, where the power reached 0.99, similar to the Cutoff method. In Scenario 7, where the prevalence of the effective subspace remained low, but the treatment effect increased from 5 to 7.5, the power of the FK method improved substantially from 0.58 (in Scenario 2) to 0.68. The FK-BMA methods were more conservative, leading to lower power in most scenarios, especially when the effective subspace prevalence was low (Scenarios 2 and 7). 

The accuracy of treatment recommendations generally mirrored the patterns observed in power across the different methods. In scenarios where the Cutoff method was correctly specified (2 and 6), it outperformed the other methods, achieving a high accuracy of 0.92. These figures highlight the method's positive performance when the underlying model assumptions are met. However, in scenarios where the Cutoff model assumptions were not met, such as Scenario 8, its accuracy dropped to 0.51. The FK method consistently achieved the highest accuracy, except for Scenarios 2 and 6 in which the Cutoff method performed best. FK's accuracy ranged from 0.66 in Scenario 8 to 0.96 in Scenario 1. The FK-BMA methods, while showing slightly lower accuracy compared to the FK method, still demonstrated reasonable performance, particularly in scenarios with moderate effective subspace prevalence. The accuracy was similar across the two $\lambda_2$ values: the median accuracies were 0.70 [interquartile range (IQR): 0.59, 0.75] and 0.70 [IQR: 0.62, 0.77] for $\lambda_2=2$ and $\lambda_2=3$, respectively.

\begin{table*}[t]
\centering
\caption{Exact detection rate (EDR; the success rate of exactly identifying the appropriate predictive variables without including any incorrect variables) and inclusive detection rate (IDR; the success rate of having the appropriate predictive variables included in the detected set, even if other variables are also included) for the FK-BMA method under eight simulation scenarios and two different model sparsity parameters.}
\label{tab:rr}
\begin{tabular}{c|cccc}
  \hline
  & \multicolumn{2}{c}{FK-BMA ($\lambda_2 = 2$)} & \multicolumn{2}{c}{FK-BMA ($\lambda_2 = 3$) } \\ 
Scenario & EDR & IDR &  EDR & IDR \\
  \hline
1 & 0.58 & 1.00 & 0.15 & 1.00 \\ 
  2 & 0.07 & 0.49 & 0.01 & 0.71 \\ 
  3 & 0.25 & 0.75 & 0.03 & 0.93 \\ 
  4 & 0.15 & 0.97 & 0.01 & 0.98 \\ 
  5 & 0.16 & 0.48 & 0.03 & 0.73 \\ 
  6 & 0.23 & 0.83 & 0.08 & 0.97 \\ 
  7 & 0.20 & 0.62 & 0.14 & 0.87 \\ 
  8 & 0.16 & 0.55 & 0.14 & 0.91 \\ 
   \hline
\end{tabular}
\end{table*}

Table \ref{tab:rr} presents the exact detection rate (EDR) and inclusive detection rate (IDR) for the FK-BMA method across eight simulation scenarios, using two different model sparsity parameters (\(\lambda_2 = 2\) and \(\lambda_2 = 3\)). The scenarios vary in the number of predictive variables, with Scenario 1 containing no predictive variables, Scenario 2 containing two, and Scenarios 3 through 8 each containing one. A clear trend emerged regarding the choice of sparsity parameter \(\lambda_2\). FK-BMA with \(\lambda_2 = 2\) generally exhibited higher EDRs compared to \(\lambda_2 = 3\), indicating that a less sparse model (\(\lambda_2 = 2\)) was more effective at exactly identifying the correct predictive variables. However, EDRs were generally low across both settings, a median of 0.18 (IQR: 0.16–0.23) for \(\lambda_2 = 2\) and a lower median of 0.06 (IQR: 0.02–0.14) for \(\lambda_2 = 3\). When considering the IDR, FK-BMA with \(\lambda_2 = 3\) generally performed better, with a median IDR of 0.92 (IQR: 0.83–0.97), compared to 0.68 (IQR: 0.53–0.87) for \(\lambda_2 = 2\). The performance was notably poor in Scenario 2, where the EDR was only 0.07 for \(\lambda_2 = 2\) and 0.01 for \(\lambda_2 = 3\) and the IDR was 0.49 for \(\lambda_2 = 2\) and 0.71 for \(\lambda_2 = 3\), reflecting difficulties in exactly identifying the two predictive variables present in this scenario. 

Overall, the inclusive detection rates improve as \(\lambda_2\) increases, with the performance of FK-BMA more closely resembling that of FK. Conversely, the exact detection rates tend to worsen as \(\lambda_2\) increases, reflecting the trade-off between inclusiveness and precision in variable selection. Our findings show inconsistent identification rates across  scenarios, with generally poor performance. This suggests that the FK-BMA method's ability to correctly identify predictive variables are  sensitive to trial-specific parameters.

\begin{table*}[t]
\centering
\caption{Expected sample size for each method under eight simulation scenarios.}
\label{tab:trial_size}
\begin{tabular}{c|cLLc}
  \hline
  & \multicolumn{4}{c}{Expected Sample Size}  \\ 
Scenario & Cutoff & FK-BMA ($\lambda_2 = 2$) & FK-BMA ($\lambda_2 = 3$) & FK \\
  \hline
1 & 450.35 & 460.04 & 460.29 & 455.62 \\ 
  2 & 387.53 & 464.00 & 461.59 & 419.25 \\ 
  3 & 357.46 & 356.51 & 353.04 & 333.52 \\ 
  4 & 313.01 & 332.43 & 328.77 & 308.15 \\ 
  5 & 314.61 & 313.81 & 312.93 & 304.65 \\ 
  6 & 309.62 & 348.81 & 344.79 & 317.77 \\ 
  7 & 437.26 & 453.58 & 449.57 & 397.25 \\ 
  8 & 390.76 & 391.74 & 389.08 & 356.38 \\ 
   \hline
\end{tabular}
\end{table*}

When comparing the four methods—FK, Cutoff, and the two FK-BMA variants (with \(\lambda_2 = 2\) and \(\lambda_2 = 3\))—distinct patterns emerge in terms of sample size efficiency (Table \ref{tab:trial_size}). Overall, FK and Cutoff consistently required fewer participants than the FK-BMA methods, particularly in more complex scenarios. This is expected as the FK-BMA methods additionally incorporate variable selection. In the null scenario (Scenario 1), all methods had similar sample sizes, with FK (455.62) and Cutoff (450.35) being slightly more efficient than FK-BMA. As the complexity increased, especially in Scenario 2 with the cutoff correctly specified and a small effective subspace prevalence (0.18), the Cutoff method demonstrated clear efficiency, requiring only 387.53 participants, compared to FK's 419.25 and the larger sample sizes demanded by FK-BMA (\(\lambda_2 = 2\) at 464.00 and \(\lambda_2 = 3\) at 461.59). In scenarios where the cutoff is incorrectly specified, FK generally required fewer participants than Cutoff, though the differences were often modest.

\begin{table*}[t]
\centering
\caption{Empirical type I error (T1E; Scenario 1), power (Scenarios 2), accuracy of treatment recommendations on an external dataset ($n=10,000$), and expected sample size for two simulation scenarios with VCB added as a third predictive variable.}
\label{tab:three_vars}
\begin{tabular}{c|cccccc}
  \hline
  & \multicolumn{2}{c}{Power}  & \multicolumn{2}{c}{Accuracy} & \multicolumn{2}{c}{Expected Sample Size} \\ 
Scenario & Cutoff & FK & Cutoff & FK & Cutoff & FK \\
  \hline
1 & 0.08 (T1E) & 0.03 (T1E) & 0.98 & 0.98 & 448.10 & 451.89 \\ 
  2 & 0.99 & 0.97 & 0.64 & 0.75 & 320.86 & 335.21 \\ 
   \hline
\end{tabular}
\end{table*}

We additionally conducted a simulation study assessing trial performance when adding a third candidate predictive variable, VCB (Table \ref{tab:three_vars}, see Web Appendix A for information on models and simulation parameters). In Scenario 1, where no treatment effect is present, the empirical type I error was controlled within acceptable limits for the FK method (0.03), while the Cutoff method exceeded the threshold, with a type I error of 0.08. Both methods demonstrated high accuracy in treatment recommendations, each achieving an accuracy rate of 0.98. The expected sample sizes were also similar, with Cutoff requiring 448.10 participants and FK requiring 451.89 participants, indicating comparable performance in the null scenario. Scenario 2 showed strong performance for both methods in terms of power, with both methods achieving high power ($\ge 0.97$). FK demonstrated slightly better accuracy (0.75) compared to Cutoff (0.64). However, the Cutoff method was more efficient, having a smaller expected sample size of 320.86 participants compared to 335.21 for FK.

{

\subsection{Illustrative Simulated Example}
\label{s:illu_sim_ex}
This section presents an illustrative simulated example demonstrating the execution and interpretation of the proposed methods in the context of a Bayesian adaptive enrichment trial. The example utilizes the \texttt{rjMCMC} function from the R package \texttt{fkbma} to assess biomarker-treatment relationships and identify potential subgroups for treatment efficacy. Specifically, we consider Scenario 4 from the simulation study, where the continuous biomarker HB exhibits a nonlinear relationship with the treatment effect. This example illustrates how the method operates in a realistic setting, highlighting the impact of flexible modeling on decision-making in clinical trials.

Data were generated for 500 patients, following the specifications of Scenario 4. Biomarkers HB and $\Delta$HR were drawn from truncated t-distributions, tuned based on their empirical distributions from previous observational studies (Section \ref{s:sim_setup}). The treatment assignment was binary, sampled from a Bernoulli distribution with a probability of 0.5. The response variable $Y_i$ was modeled as a function of HB, its interaction with treatment, and random noise:
\[
Y_i = 3I(x_{1i}\ge 60) + \left[\frac{\exp\{30(F_{\text{HB}}(x_{1i}) - 0.5)\}}{1 + \exp\{30(F_{\text{HB}}(x_{1i}) - 0.5)\}} \right] \cdot T_i + \epsilon_i,
\]
where \( F_{\text{HB}}(x_{1i}) \) is the empirical cumulative distribution function (ECDF) of HB, \( T_i \) represents the treatment indicator, and \( \epsilon_i \sim \mathcal{N}(0, 5) \) represents random noise. HR, $\Delta$HB, and their interaction were considered as candidate prognostic and predictive variables.

The rjMCMC procedure was configured to ensure reliable posterior sampling while mitigating issues related to convergence and autocorrelation. A total of 2000 posterior samples were collected after an initial burn-in period of 5000 iterations, allowing the Markov chain to stabilize before inference. To further reduce autocorrelation in the samples, thinning was applied, retaining only every fifth iteration. Multiple independent chains (four in total) were run to facilitate robust convergence diagnostics. Additionally, the proposal variance for MCMC updates was set to 0.1, controlling the step size in parameter updates.

The prior distributions were specified to be uninformative, except for \(\lambda_2\), which governs the number of knots in the model and is tuned via simulation. A weakly informative inverse gamma prior was placed on the residual variance \(\sigma_\epsilon^2\) with shape and scale parameters set to small values (both 0.01), ensuring minimal influence on the posterior distribution. The degree of V-splines used in modeling continuous variables was set to one, corresponding to piecewise linear functions, while the maximum number of candidate knots was constrained to five per spline. The prior variance on regression coefficients was set to \(\sqrt{20}\), reflecting no prior information on about the magnitude of the effects.

\begin{figure}
    \centering
    \includegraphics[width=0.8\textwidth]{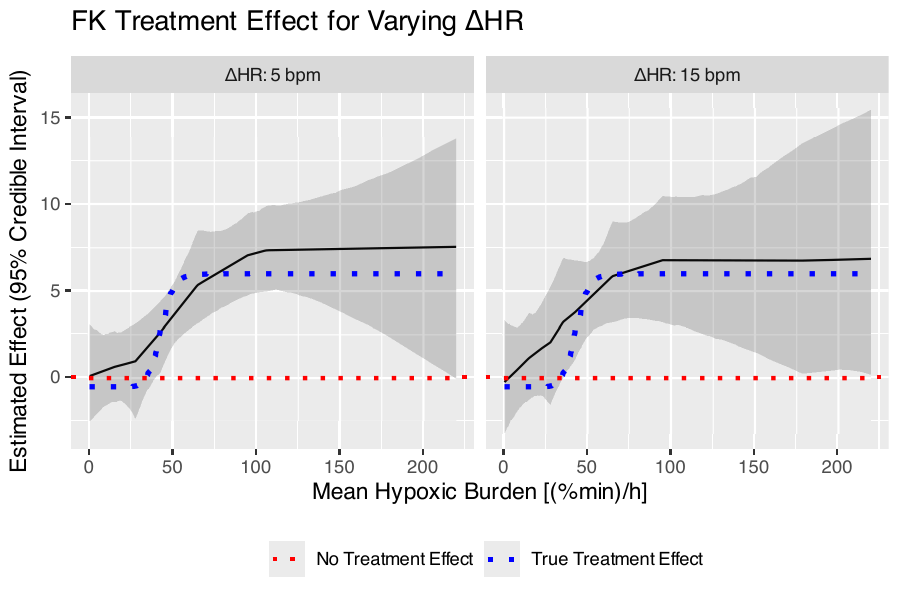} 
    \caption{Posterior mean of the treatment effect as a function of HB and $\Delta$HR. The credible intervals represent the 95\% posterior uncertainty for the estimated effect.}
    \label{fig:hr_hb_plot}
\end{figure}

Figure \ref{fig:hr_hb_plot} shows the posterior mean of the treatment effect as a function of HB, derived using the \texttt{rjMCMC} function, across two levels of $\Delta$HR (5 bpm and 15 bpm). The 95\% credible intervals illustrate the uncertainty in the estimates. The figure reveals that the relationship between HB and treatment effectiveness is consistent across different values of $\Delta$HR, indicating that $\Delta$HR does not meaningfully alter the pattern of treatment effect. Moreover, the estimated treatment pattern closely follows the true underlying pattern, showcasing the model's ability to accurately capture the relationship between HB and treatment efficacy. The estimated thresholds for treatment effectiveness, based on a significance cutoff of $\alpha=0.05$, fall within the range of 32.2–36.0 \%min/h, closely aligning with the true threshold of 31.9 \%min/h. MCMC diagnostics for the estimated treatment effects, including trace plots for representative covariate patterns, are provided in Web Appendix B.

}

\section{Discussion}

In this work, we describe a Bayesian adaptive enrichment clinical trial case study to identify subgroups where PAP treatment of OSA results in an improvement in the primary outcome measure of 24-hour systolic blood pressure under realistic sample sizes, effect sizes, and biomarker distributions. We compared three model formulations, together with identical interim analysis and decision rules: the Cutoff model, which dichotomizes continuous biomarkers HB and $\Delta$HR using predefined thresholds; the FK-BMA model, which automatically identifies optimal cutoffs and incorporates variable selection to identify the most relevant predictors; and the FK model, similar to FK-BMA, but without any variable selection. We allow for early stopping due to efficacy or futility at an interim analysis, and interim enrichment with individuals who are expected to benefit from treatment based on the currently-accumulated data.
{

This study emphasizes the real-world applicability of the proposed methods by incorporating information derived from the SHHS and MESA observational datasets. The use of actual biomarker and outcome distributions ensures that the simulations reflect realistic clinical settings, including challenges such as skewness, truncations, and variability in measurements. In contrast to theoretical evaluations in previous studies \cite{maleyeff2024adaptive}, this work applies advanced Bayesian methods to a real-world context, offering a practical case study for investigators. These findings provide a valuable framework to guide the choice between methods and to implement statistical analyses effectively in clinical practice. 

In comparing the Cutoff method and the FK method, we observed distinct strengths and limitations. The Cutoff method, while simple and interpretable, relies on pre-specified thresholds and performs well only when these cutoffs are correctly specified. This limitation makes it prone to missing complex, non-linear relationships and highlights the challenges of accurately determining cutoffs \textit{a priori}. Conversely, the FK method uses splines to dynamically model biomarker-outcome relationships, capturing nuanced patterns. However, it requires advanced computational techniques like rjMCMC for model fitting and is sensitive to the size of the effective subspace. For instance, the FK method struggled when the effective subspace was small ($\le$0.18) but achieved higher power when the effect size increased.

The FK-BMA method encountered challenges due to data limitations, as the available information was insufficient to simultaneously infer both the model structure and the parameters. Specifically, the combination of individual standard deviations, effect sizes, and sample size introduced variability that the FK-BMA method struggled to handle robustly. Consequently, it was less reliable in identifying key predictive variables essential for effective modeling. Notably, these findings differ from those reported in \cite{maleyeff2024adaptive}, where the FK-BMA method performed well under alternative trial conditions. Although the simulation study was limited to the specific example considered, we can conclude that in general, when the signal-to-noise ratio is small, BMA will not perform well, as evidenced by the results. These findings underscore the critical need for comprehensive design simulation studies to tailor model selection to the unique characteristics of each trial, ensuring the adoption of the most appropriate approach. Future work will explore a broader range of signal-to-noise ratios to more precisely determine when the performance of BMA starts to deteriorate. 
}

{
}

In sum, Bayesian adaptive enrichment designs are crucial in bridging the gap between clinical research and practical application, particularly in settings where there is evidence of treatment effect heterogeneity. For example, in OSA and related areas, existing trials have often reported conflicting results due to variability in treatment responses between different subgroups. Despite the presence of a modest number of well-defined biomarkers, their exact clinical effect is still uncertain, and the thresholds for defining treatment effect heterogeneity are not well-established. This creates specific challenges in determining which subgroups will most benefit from treatment. Bayesian adaptive designs are particularly valuable in these cases, allowing for dynamic updates to the trial population and treatment strategies based on interim data, ensuring that the study remains responsive to emerging evidence and more precisely targets effective therapies for specific patient groups. This tailored approach ensures that interventions are more effectively matched to patient needs, improving patient outcomes and advancing the integration of precision medicine into clinical practice.

\newpage

\section*{Acknowledgements}
\noindent Lara Maleyeff acknowledges the support of the Canadian Network for Statistical Training in Trials (CANSTAT) Postdoctoral Fellowship during the preparation of this research. Shirin Golchi is a Fonds de Recherche du Québec, Santé, Chercheuse-boursière (Junior 1) and acknowledges support from a Natural Sciences and Engineering Council of Canada (NSERC) Discovery Grant, Canadian Statistical Sciences Institute, and the Fonds de Recherche du Québec, Nature et technologies (FRQNT-NSERC NOVA). Erica E. M. Moodie is a Canada Research Chair (Tier 1) in Statistical Methods for Precision Medicine and acknowledges the support of a Chercheur de Mérite Career Award from the Fonds de Recherche du Québec, Santé. This work was supported by the National Institute of Mental Health of the National Institutes of Health under Award Number R01 MH114873. The content is solely the responsibility of the authors and does not necessarily represent the official views of the National Institutes of Health. R. John Kimoff was supported by Canadian Institutes of Health Research grant numbers PJT-148763 and PJT-178128 for work related to this study.

\section*{Disclosure statement}
\noindent No potential conflict of interest was reported by the authors.

\section*{Data availability statement}
\noindent The R code to implement the proposed \texttt{rjMCMC} procedure is available in the R package \texttt{fkbma}.

\bibliographystyle{unsrt}  
\bibliography{main}  
\newpage
 \section*{Web Appendix A: Simulation Study with an Additional Biomarker, Vasoconstrictive Burden}
  We conducted an additional simulation study to evaluate the performance of the the performance of the Cutoff and FK methods when incorporating a third biomarker, vasoconstrictive burden (VCB). We first let $x_{3i}$ be the VCB for individual $i$. Using a cutoff of 20 voltage$\times$min for VCB, the Cutoff model formulation is
\small
\begin{equation}
\begin{split}
    Y_i = &\beta_0 + \beta_1z_{1i} + \beta_2z_{2i} + \beta_3z_{3i} + \beta_4z_{1i}z_{2i} + \beta_5z_{1i}z_{3i} + \beta_6z_{12i}z_{3i} + \beta_7z_{1i}z_{2i}z_{3i}+\\ & \left(\beta_8 + \beta_9z_{1i} + \beta_{10}z_{2i}  + \beta_{11}z_{3i} + \beta_{12}z_{1i}z_{2i} + \beta_{13}z_{1i}z_{3i} + \beta_{14}z_{2i}z_{3i} + \beta_{15}z_{1i}z_{2i}z_{3i}\right) T_i + \epsilon_i,
\end{split}
\end{equation}
\normalsize
where $z_{1i}=I(x_{1i}>60)$, $z_{2i}=I(x_{2i}>8)$, $z_{3i}=I(x_{3i}>20)$, and $\epsilon_i \sim \mathcal{N}(0,\sigma_\epsilon^2)$. Here, $\gamma(\mathbf{x})=\beta_8 + \beta_9z_{1} + \beta_{10}z_{2}  + \beta_{11}z_{3} + \beta_{12}z_{1}z_{2} + \beta_{13}z_{1}z_{3} + \beta_{14}z_{2}z_{3} + \beta_{15}z_{1}z_{2}z_{3}$, where $\mathbf{x}=(x_{1},x_{2},x_{3})$. Similarly, the FK method is formulated as:
\small
\begin{equation}
\begin{split}
    Y_i&= \sum_{j=1}^{3} h_{j}(x_{ji}) + \sum_{1\le k \le j \le 3} h_{kj}(x_{ki}x_{ji})+ h_{123}(x_{1i}x_{2i}x_{3i}) + \\ &\left(\phi + \sum_{j=1}^{3} f_{j}(x_{ji}) + \sum_{1\le k \le j \le 3} f_{kj}(x_{ki}x_{ji}) + f_{123}(x_{1i}x_{2i}x_{3i})\right) T_i + \epsilon_i,
\end{split}
\end{equation}
\normalsize

\noindent where $h_{\cdot}$, $f_{\cdot}$ are biomarker terms represented by free-knot B-spline functions with degree 1 and $\epsilon_i \sim \mathcal{N}(0,\sigma_\epsilon^2)$. Here, the variable-specific treatment effect is $\gamma(\mathbf{x}) = \phi + \sum_{j=1}^{3} f_{j}(x_{j}) + \sum_{1\le k \le j \le 3} f_{kj}(x_{k}x_{j}) + f_{123}(x_{1}x_{2}x_{3})$. 

We use identical priors, tuning parameters, and interim analysis procedures described above with $\lambda_1$ increased to 5. Two simulation scenarios were evaluated for each design, as detailed in Table 2 of the main paper. Data were generated according to the following model:
\begin{equation}
    Y_i = 3 I(x_{1i} \ge 60) + \gamma(\mathbf{x}_i)T_i + \epsilon_i,
\end{equation}
\noindent $\epsilon_i \sim \mathcal{N}(0,8.4^2)$. VCB was generated from a truncated normal distribution $\mathcal{N}(20,13^2)$  left truncated at $0$ (Figure 1  of the main paper). We assessed two scenarios: (1) a null scenario matching Scenario 1 from the previous simulation study, and (2) a scenario where HB served as the predictive variable, with the functional form corresponding to Scenario 4. Results from the simulation study are in Table 6 of the main paper.

\section*{Web Appendix B: MCMC Convergence Diagnostics for Illustrative Simulation Example}
In this section, we describe the convergence diagnostics for the MCMC sampling procedure in the presence of free-knot splines, a setting where standard diagnostic tools may be challenging to interpret. The flexibility of free-knot splines introduces additional complexity in the posterior exploration, as the number and location of knots can vary across iterations, leading to potentially irregular mixing patterns. To assess the stability and convergence of the treatment effect estimates, we examine the MCMC trajectories for a sample of representative covariate patterns.

\begin{figure}
    \centering
    \includegraphics[width=0.8\textwidth]{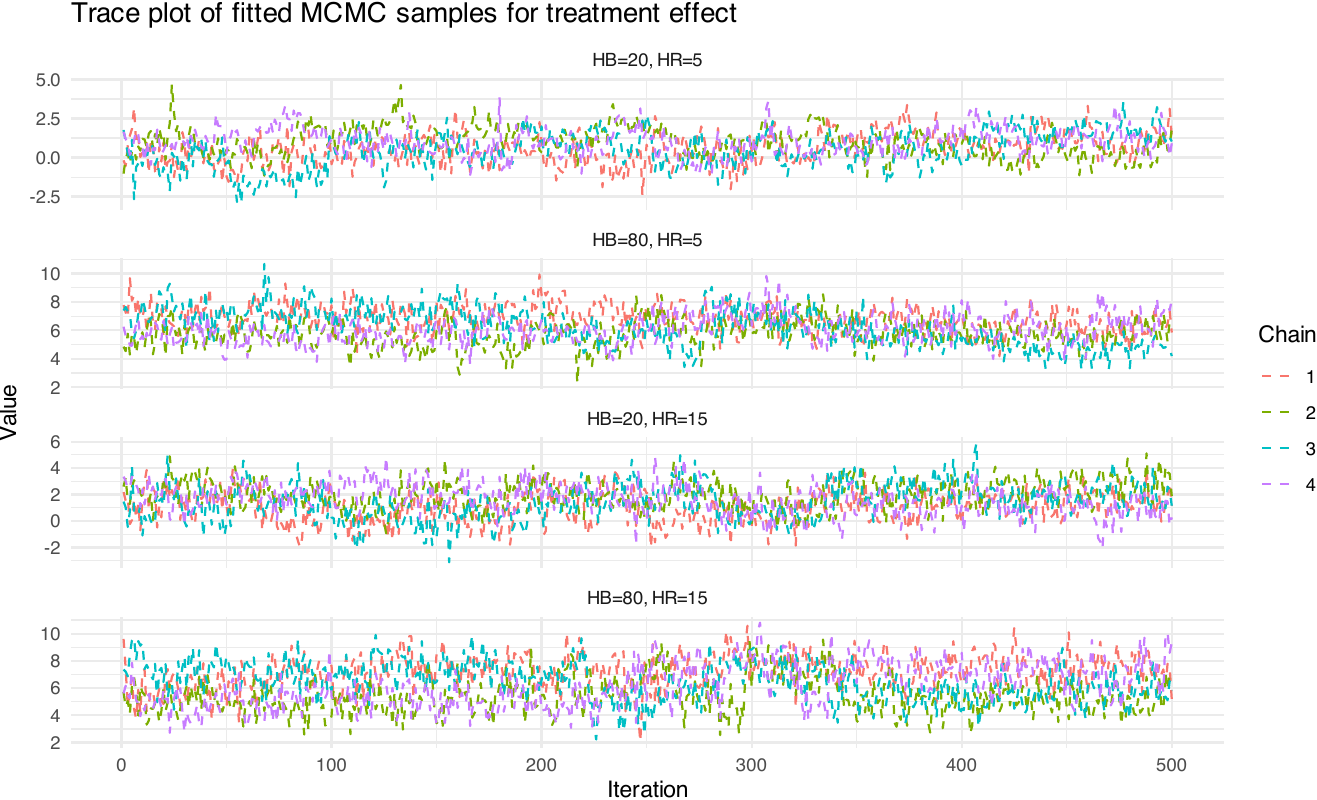}
    \caption{Trace plot showing the MCMC sampling trajectories for the treatment effect, stratified by values of HB and $\Delta$HR. Specifically, the plot includes individuals with HR set to 20 and 80, and $\Delta$HR set to 5 and 15. This plot allows visual assessment of the treatment effect's stability and convergence across varying biomarker values.}
    \label{fig:trace_plot}
\end{figure}

The trace plot (Figure~\ref{fig:trace_plot}) displays the MCMC sampling trajectories for the treatment effect, stratified by values of HB and $\Delta$HR. In particular, the plot includes individuals with HR set to 20 and 80, and $\Delta$HR set to 5 and 15. This stratification allows for a detailed visual assessment of the stability and convergence of the treatment effect across different levels of HB and $\Delta$HR. The chains demonstrate good mixing, as evidenced by their smooth trajectories across iterations without observable trends, indicating that the MCMC sampler effectively explores the posterior distribution for the treatment effect within each subgroup. 

\end{document}